\documentclass[twocolumn,superscriptaddress,showpacs,preprintnumbers,amsmath,amssymb]{revtex4}

\usepackage{graphicx}
\usepackage{dcolumn}
\usepackage{bm}

\begin{document}


\title{Quantum conductance of graphene nanoribbons with edge defects}

\author{T. C. Li}

\affiliation{Department of Physics, The University of Texas at
Austin, Austin, TX 78712, USA}
 \affiliation{Center for Nonlinear
Dynamics, The University of Texas at Austin, Austin, TX 78712, USA}
\author{Shao-Ping Lu}
\affiliation{Department of Physics, The University of Texas at
Austin, Austin, TX 78712, USA}

\date{\today}

\begin{abstract}
The conductance of metallic graphene nanoribbons (GNRs) with single
defects and weak disorder at their edges is investigated in a
tight-binding model. We find that a single edge defect will induce
quasi-localized states and consequently cause zero-conductance dips.
The center energies  and breadths of such dips are strongly
dependent on the geometry of GNRs. Armchair GNRs are much more
sensitive to a vacancy than zigzag GNRs, but are less sensitive to a
weak scatter. More importantly, we find that with a weak disorder,
zigzag GNRs will change from metallic to semiconducting due to
Anderson localization. But a weak disorder only slightly affects the
conductance of armchair GNRs. The influence of edge defects on the
conductance will decrease
 when the widths of GNRs increase.
\end{abstract}


\maketitle

\section{\label{sec1} Introduction}
Recently,  graphene (a single atomic layer of graphite) sheets were
successfully isolated for the first time  and demonstrated to be
stable under ambient conditions by Novoselov et
al.\cite{novoselov04, novoselov05} Due to their unique two
dimensional(2D) honeycomb structures, their mobile electrons behave
as massless Dirac fermions,\cite{wallace47,novoselov05,zhang05}
making graphene an important system for fundamental
physics\cite{gusynin05,peres06,aleiner06}. Moreover, graphene sheets
have the potential to be sliced or lithographed to a lot of
patterned graphene nanoribbons (GNRs)\cite{banerjee05} to make
large-scale integrated circuits\cite{wilson06}.

The electronic property of GNRs has attracted increasing attention.
Recent studies using tight-binding models\cite{nakada96,ezawa06} and
the Dirac equation\cite{brey06} have shown that GNRs can be either
metallic or semiconducting, depending on their shapes. This allows
GNRs to be used as both connections and functional
elements\cite{wakabayashi00, obradovic06} in nanodevices, which is
similar to carbon nanotubes(CNTs).\cite{yao99,chen06}

\begin{figure}[bhtp]
\setlength{\unitlength}{1cm}
\begin{picture}(8,6.5)
\put(0,6.4){(a) Zigzag ribbon}
\put(0.8,3.6){\includegraphics[totalheight=2.7cm]{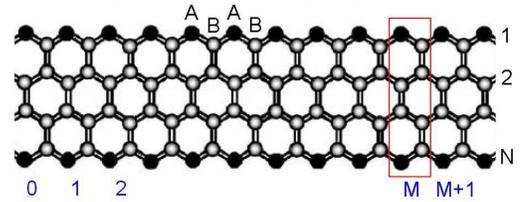}}
\put(0,2.9){(b) Armchair ribbon}
\put(1,0.0){\includegraphics[totalheight=2.7cm]{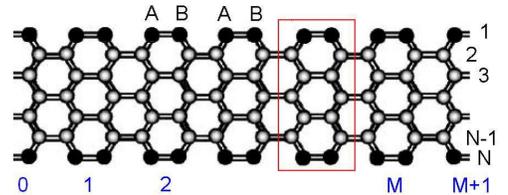}}
\end{picture}
\caption{\label{fig1} (Color online) Geometry of graphene ribbons.
(a): A zigzag ribbon (N=4); (b): An armchair ribbon (N=7). A black
circle denotes an edge carbon  and a gray circle denotes a bulk
carbon. A unit cell contains $2N$ atoms. From the top down, atoms in
a unit cell are labeled as $1$A, $1$B, $2$A, $2$B, ..., $N$A, $N$B.
Atoms close to 1B are 1A and 2A, and so on.  }
\end{figure}

However, GNRs are substantially different from CNTs by having two
open edges at both sides (see Fig.\ref{fig1}).   These  edges not
only remove the periodic boundary condition along the circumference
of CNTs, but also make GNRs more vulnerable to defects than
CNTs\cite{chico96B,anantram98}. In fact, nearly all observed
graphene edges\cite{kobayashi06,banerjee05,niimi06} contain local
defects or extended disorders, while few defects are found in the
bulk of graphene sheets. These edge defects can significantly affect
the electronic properties of GNRs. Recent theoretical studies of
perfect GNRs have also considered some
 edge corrections.\cite{ezawa06,fujita97,miyamoto99} But all edge atoms
 (see Fig. \ref{fig1}) are assumed to be identical and the
 GNRs still have translational symmetry along their axis.
There have been no studies of changes in the conductance caused by
local defects or extended disorder on the edges, which break the
translational symmetry of GNRs. We address this issue by calculating
the conductance of metallic GNRs with such edge defects  using a
tight-binding model.

In our calculation, external electrodes and the central part
(sample) are assumed to be made of GNRs. And the edge defects are
modeled by appropriate on-site (diagonal) energy in the Hamiltonian
of the sample. We utilize a quick iterative
scheme\cite{li05,sancho84} to calculate the surface Green's
functions of electrodes and an efficient recursive
algorithm\cite{li05,krompiewski02} to calculate the total Green's
function of the whole system. Finally, the conductance is calculated
by the Landauer formula.\cite{imry99,datta95} The calculation time
of this method is only linearly dependent on the length of the
sample and a GNR with disorder distributed over a length of  1
$\mu$m is tractable.

We first study the conductance of zigzag and armchair GNRs with the
simplest possible edge defects, a single vacancy or a weak scatter.
Then we use a simple 1D model to explain the zero-conductance dips
caused by edge defects.  Finally, we study some more realistic
structures, GNRs with weak scatters randomly distributed on their
edges.  We find that a weak disorder can change zigzag ribbons from
metallic to semiconducting, but only changes the conductance of
armchair ribbons slightly. The paper is organized as follows: in
section \ref{sec2}, we introduce the model and method employed in
this paper. Results and discussion are presented in section
\ref{sec3}. We conclude our findings in section \ref{sec4}.

\section{\label{sec2} Model and method}

The geometry of GNRs is shown in Fig.\ref{fig1}. A graphene ribbon
contains two unequal sublattices, denoted by A and B in this paper.
We use $N$, the number of A(B)-site atoms in a unit cell, to denote
GNRs with different widths. \cite{nakada96} Then the widths of
ribbons with zigzag edges and armchair edges are
$W_z=\frac{N}{2}\sqrt{3}\; a_0 $ and $W_a=\frac{N}{2} a_0$
respectively, where $a_0=2.49$ \r{A} is the graphene lattice
constant. From the top down, atoms in a unit cell will be labeled as
$1$A, $1$B, ..., $N$A, $N$B. As shown in Fig.
 \ref{fig1}, an edge atom is a carbon atom at the edge of GNRs that
is connected by only two other carbon atoms. In this paper, we
consider defects that  locate  at the sites of these edge atoms
only.

 The system under consideration is composed of two
electrodes and a central part(sample). The sample(unit cells 1, ...
, $M$) is a finite GNR, which may contain edge defects, while the
left and right electrodes are assumed to be semi-infinite perfect
GNRs. We describe the GNR by a tight-binding model with one
$\pi$-electron per atom. The tight-binding Hamiltonian of the system
is
\begin{equation}
\label{eq1}
H=\sum_{i}\varepsilon_i\;a_i^{\dag}\;a_i-V_{pp\pi}\sum_{<i,
j>}a_i^{\dag}\;a_j+c.c.
\end{equation}
where $\varepsilon_i$ is the on-site energy and $V_{pp\pi}$ is the
hopping parameter. The sum in $<i, j>$ is restricted  to the
nearest-neighbor atoms. In the absence of defects, $\varepsilon_i$
is taken to be zero  and $V_{pp\pi}=2.66$ eV. \cite{chico96B} In the
presence of defects, both the on-site energy and the hopping
parameter can change. Here, we only consider the variation in the
on-site energy. A vacancy is simulated by setting its on-site energy
to infinity\cite{chico96B,wakabayashi02}. A weak scatter caused by
impurity or distortion will be simulated by setting $\varepsilon_i$
to a small value $V_i$. In the case of a weak disorder, $V_i$ is
randomly selected from the interval $\pm |V_{random}|$ for every
edge atom.

In what follows we show how to calculate the conductance of the
GNRs:

First, the surface retarded Green's functions of the left and right
leads ($g^L_{0,0}$, $g^R_{M+1,M+1}$) are calculated
by:\cite{li05,jiang03,sancho84}
\begin{eqnarray}
\label{eq2}
&&g^L_{0,0}=[E^+I-H_{0,0}-H_{-1,0}^{\dag}\tilde\Lambda]^{-1},\\
\label{eq3} &&g^R_{M+1,M+1}=[E^+I-H_{0,0}-H_{-1,0}\Lambda]^{-1}
\end{eqnarray}
where $E^+$=$E+i\eta$ ($\eta \to 0^+$)\cite{eta} and $I$ is a unit
matrix. $H_{0,0}$ is the Hamiltonian of a unit cell in the lead, and
$H_{-1,0}$ is the coupling matrix between two neighbor unit cells in
the lead. Here $\Lambda$ and $\tilde\Lambda$ are the appropriate
transfer matrices, which can be  calculated from the Hamiltonian
matrix elements via an iterative
procedure:\cite{sancho84,nardelli99}
\begin{eqnarray}
\label{eq4}
\Lambda=&&t_0+\tilde{t}_0t_1+\tilde{t}_0\tilde{t}_1t_2+\ldots+\tilde{t}_0\tilde{t}_1\tilde{t}_{2}\cdots t_n,\\
\label{eq5}
\tilde\Lambda=&&t_0+t_0\tilde{t}_1+t_0t_1\tilde{t}_2+\ldots+t_0t_1t_{2}\cdots\tilde{t}_n,
\end{eqnarray}
where $t_i$ and $\tilde{t_i}$ are defined via the recursion formulas
\begin{eqnarray}
\label{eq6}
&&t_i=(I-t_{i-1}\tilde{t}_{i-1}-\tilde{t}_{i-1}t_{i-1})^{-1} t_{i-1}^2,\\
\label{eq7}
&&\tilde{t}_i=(I-t_{i-1}\tilde{t}_{i-1}-\tilde{t}_{i-1}t_{i-1})^{-1}
\tilde{t}_{i-1}^2,
\end{eqnarray}
and
\begin{eqnarray}
\label{eq8}
&&t_0=(E^+I-H_{0,0})^{-1}H_{-1,0}^{\dag},\\
\label{eq9} &&\tilde{t}_0=(E^+I-H_{0,0})^{-1}H_{-1,0}.
\end{eqnarray}
The process is repeated until $t_n$,$\tilde{t}_n\leq\delta$ with
$\delta$ arbitrarily small.

Second, including the sample as a part of the right lead layer by
layer (from $l=M$ to $l=2$), the new surface Green's functions are
found by:\cite{li05,krompiewski02}
\begin{equation}
\label{eq10}
g^R_{l,l}=[E^+I-H_{l,l}-H_{l,l+1}\,g^R_{l+1,l+1}H_{l,l+1}^{\dag}]^{-1}.
\end{equation}

Third, the total Green's function $G_{1,1}$ can then be calculated
by
\begin{equation}
\label{eq11} G_{11}=[E^+I-H_{1,1}-\Sigma^L-\Sigma^R]^{-1},
\end{equation}
where
\begin{eqnarray}
\label{eq12}
&&\Sigma^L=H_{0,1}^{\dag}g^L_{0,0}H_{0,1}\\
\label{eq13} &&\Sigma^R=H_{1,2}g^R_{2,2}H_{1,2}^{\dag}
\end{eqnarray}
are the self energy functions due to the interaction with the left
and right sides of the structure. From Green's function, the local
density of states (LDOS) at site $j$ can be found
\begin{eqnarray}
\label{eq14} &&n_j=-\frac{1}{\pi} \text{Im}[G_{(j,j)}],
\end{eqnarray}
where $G_{(j,j)}$ is the matrix element of Green's function at site
$j$.

Finally,  the conductance $G_{(E)}$ of the graphene ribbon can be
calculated  using the Landauer formula\cite{imry99,datta95}
\begin{eqnarray}
\label{eq15} &&G_{(E)}=\frac{2e^2}{h}T_{(E)}.
\end{eqnarray}
Here $T_{(E)}$ is the transmission coefficient, which can be
expressed as:\cite{fisher81,meir92}
\begin{eqnarray}
\label{eq16}
&&T_{(E)}=\text{Tr}[\Gamma^LG_{11}\Gamma^RG_{11}^{\dag}]
\end{eqnarray}
where
\begin{eqnarray}
\label{eq17} &&\Gamma^{L,R}=i[\Sigma^{L,R}-(\Sigma^{L,R})^{\dag}].
\end{eqnarray}

 In this calculation,  no matrix larger than $N \times N$ is
 involved. And its cost is only linearly dependent on the length of
 the GNRs. We have used this method to study the effects of dangling
 ends on the conductance of side-contacted CNTs.\cite{li05}
 We have also calculated the band structures of perfect GNRs by diagonalizing the Hamiltonian.
 The  conductances of
perfect GNRs agree with the band structures.

\section{\label{sec3} Results and discussion}

The electronic properties of GNRs are strongly dependent on their
geometry. There are two basic shapes of regular graphene edges,
namely zigzag and armchair edges, depending on the cutting direction
of the graphene sheet (see Fig. \ref{fig1}). All ribbons with zigzag
edges (zigzag ribbons) are metallic; however, two thirds of ribbons
with armchair edges (armchair ribbons) are
semiconducting.\cite{nakada96,ezawa06}. The bands of zigzag GNRs are
partially flat around Fermi energy ($E_F=0$ eV),\cite{nakada96}
which means the group velocity of mobile electrons is close to zero.
 On the other hand, the bands of metallic armchair GNRs
 are linear around Fermi energy.\cite{nakada96,kobayashi06} So the group
 velocity of their mobile electrons around
 Fermi energy should be a constant value, which
 is measured to be about
 $10^6 \text{m/s}$.\cite{novoselov05}  Since zigzag GNRs and armchair GNRs have
 so different electronic properties, the effects of edge defects on
 their conductance should  also be very different.

\subsection{\label{sec3:sec1} Single defects}
In this section, we will study the conductance and local density of
states(LDOS) of GNRs with some single edge defects. A study of the
effects of a single defect is not only realistic( e.g., a single
two-atom vacancy at an armchair edge has been
observed\cite{kobayashi06}), but also can serve as a guide for us to
understand the effects of more complex edge defects.

\subsubsection{\label{sec3:sec1:sec1} Single vacancies}

\begin{figure}[bhtp]
\includegraphics[totalheight=8cm]{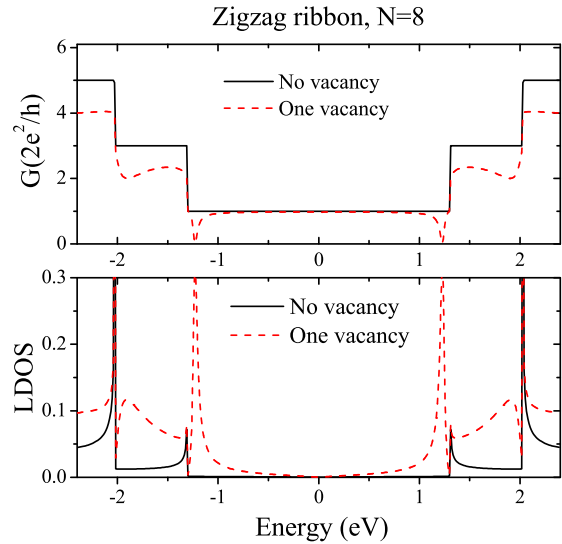}
 \caption{\label{fig2}
(Color online) (a) Conductance of a zigzag ribbon (N=8, $W=17.3$
\r{A}) with (dashed line) and without a vacancy at its edge (solid
line). (b) LDOS of a 1B atom  when there is no vacancy (solid line)
and when there is a vacancy on its nearest 1A site at the edge
(dashed line). }
\end{figure}

\begin{figure}[thp]
\includegraphics[totalheight=9cm]{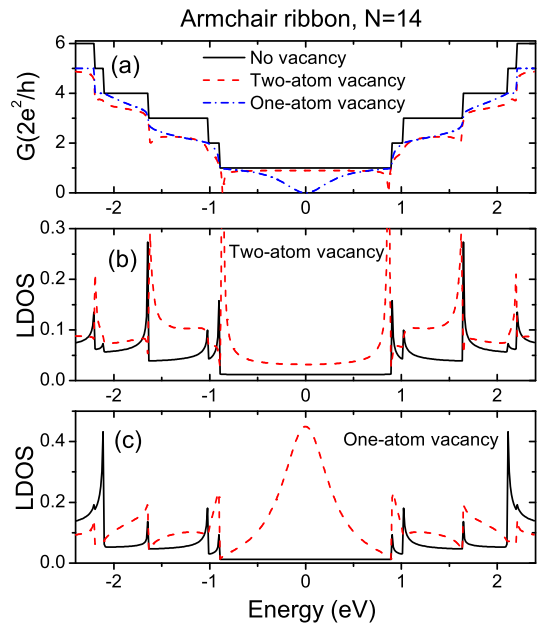}
 \caption{\label{fig3}
(Color online)
 (a) Conductance of an armchair  ribbon ($N=14$, $W=17.4$
\r{A})
 without vacancy (solid line),
 with a two-atom vacancy (dashed line) and
 with a one-atom vacancy (dash dot line) at its edge .
 (b) LDOS of a  2B atom when there is no vacancy (solid line) and when
 its nearest (1A, 1B) pair atoms are
removed (dashed line). (c) LDOS of a  1B atom when there is no
vacancy (solid line) and when its nearest 1A atom is removed (dashed
line).}
\end{figure}

One of the simplest defects in a zigzag GNR is a single vacancy
caused
 by the loss of one edge atom (one-atom vacancy).
 In Fig.\ref{fig2}(a), we plot the conductance of a zigzag GNR (N=8)
 with a single 1A vacancy (dashed line) as a function of the energy. The solid line is for
the perfect GNR. The defect almost does not affect the conductance
around the Fermi energy. There are two conductance dips close to the
first band edges and simultaneously two peaks appear in the LDOS of
the 1B atom near the vacancy (dashed line in Fig.\ref{fig2}(b)).
These two peaks have energies different from Van Hove singularities
of a perfect GNR, which are extreme points of the 1D energy bands.
So they are quasi-localized states caused by the vacancy. And the
conductance dips are due to the antiresonance of these
quasi-localized states. The relation between the conductance and
quasi-localized states will be discussed further in Sec.
\ref{sec3:sec1:sec2}.

The conductance and LDOS of armchair GNRs are displayed in
Fig.\ref{fig3}. In an armchair GNR, edge atoms appear in pairs. So a
``single vacancy" can be formed by the loss of one edge atom
(one-atom vacancy) or a pair of nearest edge atoms (two-atom
vacancy).  These two types of vacancies have very different
properties especially around the fermi energy. For the two-atom
vacancy, the conductance is  similar to that of the single vacancy
situation of the zigzag ribbon, as well as the LDOS. However, when
there is a one-atom vacancy, a large LDOS will be formed and
consequently a large conductance dip will appear at the fermi
energy. In fact, it is expected that the effect of a one-atom
vacancy is much larger than a two-atom vacancy because a one-atom
vacancy breaks the symmetry between  the two sublattices, while a
two-atom vacancy keeps such symmetry. This is the same as in CNTs.

\begin{figure}[bhtp]
\includegraphics[totalheight=6cm]{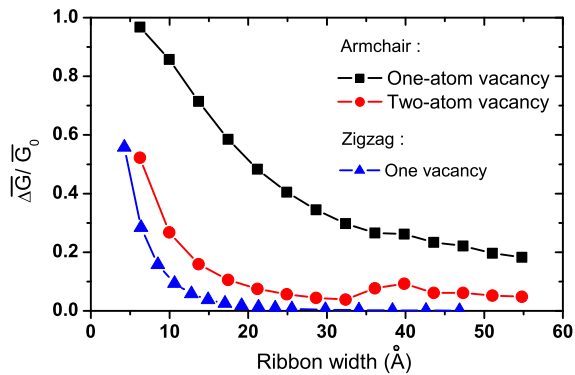}
 \caption{\label{fig4}
(Color online) The decreasing rate of the average conductance
(between $\pm 0.5$eV) due to a single vacancy at the edge. As the
width of a zigzag ribbon increases, its conductance becomes immune
from a vacancy at its edge. }
\end{figure}

 The conductance around the Fermi energy is a very important
  parameter for the application of GNRs.  It is affected by
edge defects shown above. Also it depends on the width of the
ribbon. For example, a one-atom vacancy at the edge of an armchair
GNR will always cause a zero-conductance dip at the Fermi energy.
But the breadth of the dip will change when the width of GNR
changes. In order to describe the effect of a defect on the
conductance quantitatively, we introduce the decreasing rate of the
average conductance, which is
\begin{eqnarray}
\label{eq18} {\overline{\Delta G}/\, \overline{G_0}}={\int^{\Delta
E}_{-\Delta E}[G_0(E)-G(E)] \text{d}E \over \int^{\Delta E}_{-\Delta
E}G_0(E) \text{d}E},
\end{eqnarray}
where $G_0(E)$ is the conductance of a GNR without defects, and
$G(E)$ is the conductance of the GNR with a defect. The decreasing
rate of the average conductance (between $\pm 0.5$ eV) as a function
of the ribbon width is plotted in Fig.\ref{fig4}. From the figure,
we know that edge vacancies affect armchair GNRs much more strongly
than zigzag GNRs. And the effect of edge vacancies decreases when
the width of GNRs increases. This is because there are more atoms in
the cross section of a wider GNR, so electrons are easier to go
around the defect. Thus we can use wide GNRs as connections in a
nanodevice to avoid the change of conductance due to edge vacancies.
There is a small bump in $\overline{\Delta G}/\, \overline{G_0}$ of
armchair GNRs with a two-atom vacancy at $40$\r{A}. It is because
the conductance dips at band edges(Fig. \ref{fig3}) enter into the
energy range of $\pm 0.5$ eV as the width of GNRs increases . This
does not change the overall decreasing tendency of $\overline{\Delta
G}/\, \overline{G_0}$ when the width of GNRs increases.

\subsubsection{\label{sec3:sec1:sec2} Single weak scatters}

Another kind of  single defect is a weak scatter, which can be
caused by a local lattice distortion, an absorption of an impurity
atom at the edge, or a substitution of a carbon atom by an impurity
atom. Such a single weak scatter will be simulated by changing the
on-site energy of an edge atom to a small defect potential $V$.

\begin{figure}[bhtp]
\includegraphics[totalheight=7.5cm]{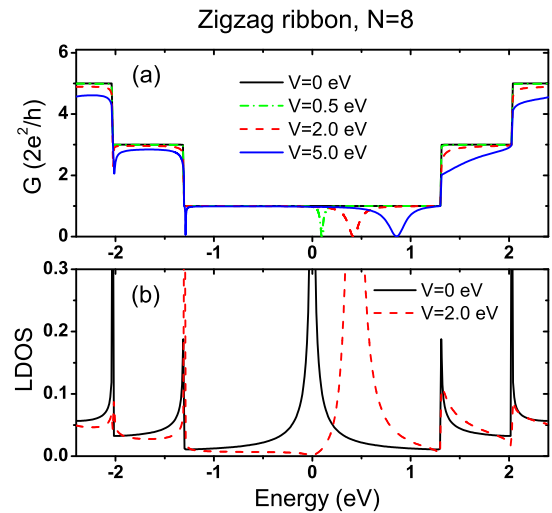}
 \caption{\label{fig5}
(Color online) (a) Conductance of a zigzag ribbon ($N=8$)
 for  various strengths of defect potential when the ribbon has a
weak scatter at its edge.
 (b) The LDOS of an edge atom for $V=0$
eV (no defect)  and $V=2.0$ eV at that site.   }
\end{figure}

\begin{figure}[bhtp]
\includegraphics[totalheight=7.5cm]{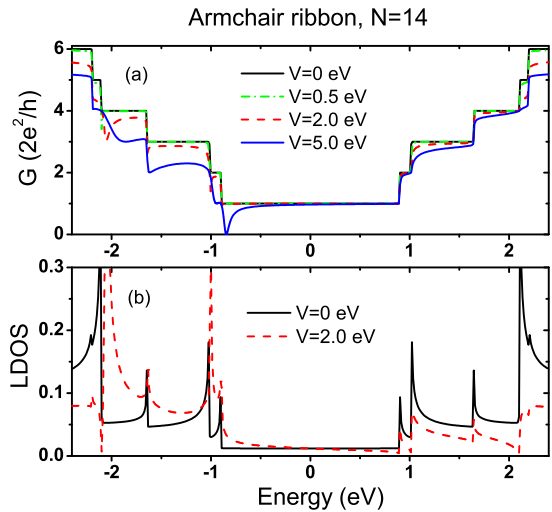}
 \caption{\label{fig6}
(Color online) (a) Conductance of an armchair ribbon ($N=14$)
 for  various strengths of defect potential when the ribbon has a
weak scatter at its edge. (b) The LDOS of an edge atom for $V=0$ eV
(no defect)  and $V=2.0$ eV at that site.  }
\end{figure}

The conductances and LDOS of  zigzag GNRs under the influences of
single weak scatters  with different strengths are presented in
Fig.\ref{fig5}. It can be seen that even a very weak scatter($V=0.5$
eV) can produce a quasi-localized state  around Fermi energy and
cause a zero-conductance dip. And the energy level and breadth of
the dip increase when the defect potential increases.  This is
because the kinetic energy of mobile electrons in a zigzag GNR is
nearly zero around Fermi energy. And these mobile electrons are
localized to ribbon edges (edge states).\cite{nakada96} So they can
be easily reflected by a weak scatter at the edge. On the other
hand,  the group velocity of mobile electrons in armchair GNRs
around Fermi energy is in the order of $10^6 \text{m/s}$, which
gives a large kinetic energy. So these mobile electrons will not as
sensitive to a weak scatter  as those in zigzag GNRs. In fact, there
is no conductance dips near Fermi energy for armchair GNRs with a
weak scatter. And the conductance dip at the band edge only becomes
visible when the defect potential is larger than $2.0 $ eV (see
Fig.\ref{fig6}).

\subsubsection{\label{sec3:sec1:sec2} A simple one dimensional model }

There are some common characters in the conductance curves and LDOS
curves shown above (Figs. \ref{fig2}, \ref{fig3}, \ref{fig5},
\ref{fig6}). First, there are sharp peaks in LDOS curves of perfect
GNRs, which are Van Hove singularities(VHS) corresponding to extreme
points in the energy bands. VHS are characteristic of the dimension
of a system. In 3D systems, VHS are kinks due to the change in the
degeneracy of the available phase space, while in 2D systems,  the
VHS appear as stepwise discontinuities with increasing
energy.\cite{odom00} Unique to 1D systems, the VHS display as peaks.
So GNRs are expected to exhibit sharp peaks in the LDOS due to the
1D nature of their band structures. Second, besides these VHS, there
are new peaks in the LDOS of GNRs with an edge defect. And
zero-conductance dips occur at the same energy of these new peaks
simultaneously. These new peaks in LDOS only occur near the defect,
but have effects on the conductance of GNRs. So they correspond to
quasi-localized states.  And the zero-conductance dips are due to
the anti-resonance of these quasi-localized states.  The relation
between quasi-localized states and zero-conductance dips can be
understood by a simple 1D model.

\begin{figure}[bhtp]
\includegraphics[totalheight=3cm]{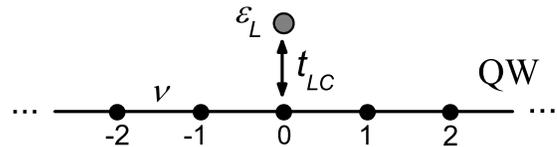}
 \caption{\label{fig7}
 A one-dimensional model of a system including conducting bands and
a quasi-localized state (QLS). The bulk of the system is represented
by a quantum wire (QW), which has one conducting band.
  }
\end{figure}

A GNR with an edge defect which induces a quasi-localized state(QLS)
can be modeled as a 1D quantum wire (QW) with a side quantum dot
(see Fig.\ref{fig7}).  The quantum wire has one conducting band with
dispersion relation $E=2\, \nu \, \cos(k d)$, where $E$ the energy
of electrons, $\nu$ the hopping coefficient in the QW and $d$ the
lattice spacing. The energy level of the quasi-localized state (side
quantum dot) is $\varepsilon_{L}$. And the coupling between the
quasi-localized state and the QW is $t_{LC}$. If $t_{LC}=0$, the
state is completely localized and has no effect on the conductance
of the QW. When $t_{LC}\neq 0$, the electrons not only can transport
in the QW, but also can transport through ``QW$\rightarrow$QLS$
\rightarrow$QW", ``QW$\rightarrow$QLS$\rightarrow$QW$\rightarrow$
QLS$\rightarrow$QW", and so on. These different channels will
interfere with each other and can cause resonance or antiresonance.
It is easy to show that they will always cause
antiresonance:\cite{orel03}

To calculate the conductance of this simple system, we assume that
the electrons are described by a plane wave incident from the far
left with unity amplitude and a reflection amplitude $r$ and at the
far right by a transmission amplitude $t$. So the probability
amplitude to find the electron in the site $j$ of the QW in the
state $k$ can be written as
\begin{eqnarray}
\label{eq19} &&a^k_j=e^{ikdj}+r\, e^{-ikdj}, \; j<0, \\
\label{eq20} &&a^k_j=t\, e^{ikdj}, \; j>0.
\end{eqnarray}
Then the transmission amplitude $t$ and thus the conductance of the
system can be easily calculated by its tight-binding
Hamiltonian.\cite{orel03} The conductance is
\begin{eqnarray}
\label{eq21} &&G_{(E)}=\frac{2e^2}{h}\frac{1}{1+\displaystyle
 \frac{t^4_{LC}}{ 4\nu^2\sin^2(kd)\, (E-\varepsilon_{L})^2}}.
\end{eqnarray}

From Eq.\ref{eq21}, we can see that when $E=\varepsilon_L$, the
conductance $G$ will be zero and a  dip will appear in the
conductance curve. In other words, the incident electrons will be
totally reflected when their energy is equal to the energy level of
the quasi-localized state. So the quasi-localized state causes an
antiresonance. This analytical result agrees with our numerical
results of GNRs. This relation between the conductance dips and
localized states is very useful in experiments. It's not easy to
measure the conductance of GNRs directly because of their small
size. But the quasi-localized states can be find in the STS
(Scanning Tunneling Spectroscopy) images or low bias STM
images.\cite{kobayashi06} Then with a STS image or a low bias STM
image, the conductance dips can be predicted.

\subsection{\label{sec3:sec2} Weak disorders}

Experimental  observed graphene
edges\cite{kobayashi06,banerjee05,niimi06} have a lot of randomly
distributed defects. Most of these defects are likely to be avoided
in future with improvements in the processing of GNRs. However, as
all materials have defects, real GNRs will always have some
uncontrollable defects at their edges due to lattice distortion or
impurity. In this section, we will consider the properties of GNRs
under the influence of weak uniform disorders at their edges. An
 edge disorder distributed over a length $L$, will be simulated
by setting the on-site energies of all edge atoms within a length
$L$ to energies randomly selected from the interval $\pm
|V_{random}|$, where $ |V_{random}|$ is the disorder strength.

\begin{figure}[bhtp]
\includegraphics[totalheight=8cm]{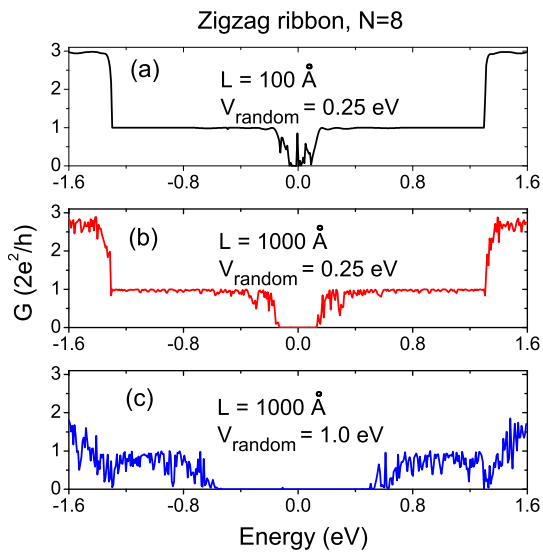}
 \caption{\label{fig8}
(Color online)  Conductance versus energy for a zigzag ribbon
($N=8$) with disorder distributed at both edges over a length of
$100$ \r{A}   and $1000$ \r{A}. With a weak disorder, zigzag GNRs
change from metallic to semiconducting.}
\end{figure}

 The conductances of  zigzag GNRs with different
disorder strengths and distribution lengths are displayed in
Fig.\ref{fig8}. The most important feature of the conductance curves
is that there are gaps around the Fermi energy. For a $N=8$
 zigzag GNR with a very weak disorder
($V_{random}=0.25$ eV) distributed over a length of 1000\r{A}, the
conductance have a $0.25$ eV gap, within which its maximum  is less
than $10^{-3}\cdot (2e^2/h)$. If the disorder strength is $1.0$ eV,
the conductance gap is $1.04$ eV. This is enormous, because a
semiconducting perfect  GNR with a similar width only has a gap less
than $0.7$ eV.\cite{ezawa06,brey06}

\begin{figure}[bhtp]
\includegraphics[totalheight=6cm]{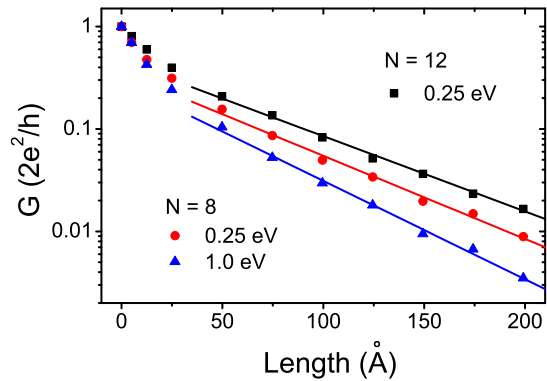}
 \caption{\label{fig9}
(Color online)  Conductance versus ribbon length for two zigzag
ribbons ($N=8$ and $N=12$) with different disorder strengths
($V_{random}=0.25$ eV and $V_{random}=1.0$ eV). The straight lines
are exponential fits to the simulated data with the ribbon length
larger than $50$ \r{A}. }
\end{figure}

The conductance gaps around the Fermi energy come from the Anderson
localization of electrons.\cite{anderson58,biel05,gornyi05} In a
perfect GNR or a GNR with periodic defects, the constructive
interference of tunneling allows that electrons within certain
energy bands can propagate through an infinite GNR (Bloch
tunneling). However, the disorder can disturb the constructive
interference sufficiently to localize  electrons. In an infinite 1D
system, even weak disorder localizes all states, yielding zero
conductance. If the disorder is only distributed within a finite
length $L$, the conductance is expected to decrease exponentially
with length, $G=G_0 \exp(-L/L_0)$, when $L$ is much larger than
localization length $L_0$.\cite{lee85,beenakker97} We observe this
is true in our simulations (see Fig. \ref{fig9}). Each point in Fig.
\ref{fig9} is an average over several thousand disorder
configurations. The localization length of electrons with energy
close to zero is very small in the zigzag ribbons. From the top
down, the fitted localization lengths are $L_0=59$\r{A},
$L_0=53$\r{A}  and $L_0=45$\r{A} for curves in Fig.\ref{fig9},
respectively. So  the wider the ribbon, the longer the localization
length. And the stronger the disorder strength, the shorter the
localization length.

\begin{figure}[bhtp]
\includegraphics[totalheight=6cm]{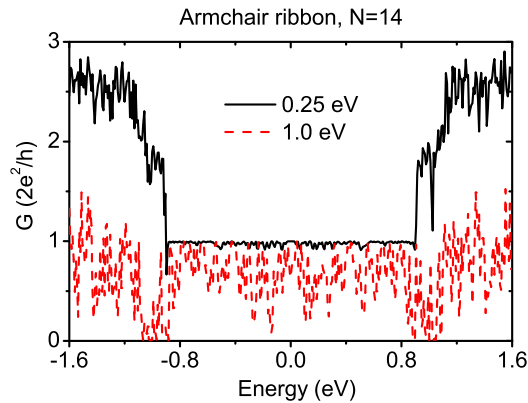}
 \caption{\label{fig10}
(Color online) Conductance versus energy for an armchair ribbon
($N=14$) with disorder distributed at its both edges over a length
of 1000 \r{A}.  }
\end{figure}

\begin{figure}[bhtp]
\includegraphics[totalheight=6cm]{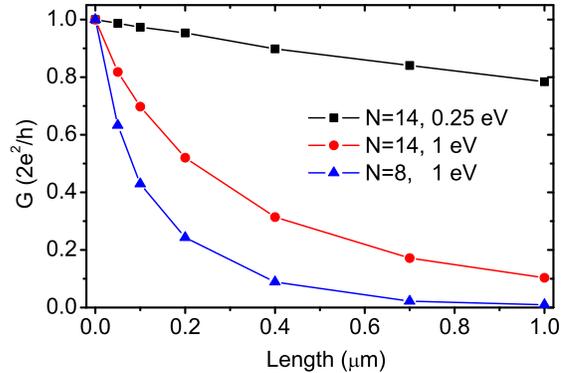}
 \caption{\label{fig11}
(Color online)  Conductance versus ribbon length for two armchair
ribbons ($N=8$ and $N=14$) with different disorder strengths.}
\end{figure}

The conductance of the armchair GNRs with weak disorders  is plotted
in Fig.\ref{fig10}. Unlike the conductance of zigzag ribbons, there
is no gap around the Fermi energy.   We also calculate the
conductance versus disorder length, which is shown in
Fig.\ref{fig11}. The conductance also decreases exponentially but
much slower. The $N=14$
 armchair GNRs ($W=17.4$\r{A}) have nearly the same width of
$N=8$ zigzag GNRs ($W=17.3$\r{A}). But their localization lengths
are very different. When the disorder strength is $V_{random}=0.25$
eV, the localization length of $N=14$
 armchair GNRs is larger than $2$ $ \mu$m, while
$L_0=59$\r{A} in $N=8$ zigzag GNRs. So the localization is much
weaker in armchair GNRs than in zigzag GNRs. As discussed in Sec.
\ref{sec3:sec1}, this is because the kinetic energy of mobile
electrons around Fermi energy in armchair GNRs are larger than in
zigzag GNRs, and also because these mobile electrons in zigzag GNRs
are localized to edges, while they distribute in the whole cross
section of armchair GNRs. So when compared to zigzag GNRs, armchair
GNRs are  more like  2D systems where electrons are easier to travel
around defects. There is no such difference between zigzag and
armchair CNTs. The conductances of both zigzag CNTs and armchair
CNTs are not significantly affected by disorder.\cite{anantram98}

Recent studies of perfect GNRs have found that, with edge
corrections which keep the translational symmetry of GNRs,  all
zigzag GNRs will be still metallic.\cite{ezawa06} Here we shown that
with a weak disorder at edges, zigzag GNRs will change from metallic
to semiconducting due to Anderson localization. So narrow zigzag
GNRs with a weak disorder can be used as functional elements  in a
nanodevice. This result is important because nearly all realistic
GNRs contain some edge disorder.

\section{\label{sec4} Conclusion}

Using a tight-binding model, we have investigated the conductance of
the zigzag and armchair graphene nanoribbons with single defects or
weak disorder at their edges. We first study the simplest possible
edge defects, a single vacancy or a weak scatter. We find that even
 these simplest defects have highly  non-trivial effects. A single edge
defect will induce quasi-localized states and consequently cause
zero-conductance dips. And the center energies and breadths of such
dips are strongly dependent on the geometry of GNRs. A one-atom edge
vacancy will completely reflect electrons at Fermi energy in an
armchair GNR, while only slightly affecting the transport of
electrons in a zigzag GNR. The effect of a two-atom vacancy in an
armchair GNR is similar to the effect of  a one-atom vacancy in a
zigzag GNR. A weak scatter can cause a quasi-localized state and
consequently a zero-conductance dip near Fermi energy in a zigzag
GNR. But its effect on the conductance of armchair ribbons near
Fermi energy is negligible. The influence of edge defects on the
conductance will decrease when the widths of GNRs increase. Then we
use a simple one dimensional model to discuss the relation between
quasi-localized states and zero-conductance dips of GNRs. We find
that a quasi-localized state caused by a defect will cause
antiresonance and corresponds to a zero-conductance dip.

Finally, we study some more realistic structures, GNRs with weak
scatters randomly distributed on their edges. We find that with a
weak disorder distributed in a finite length, zigzag GNRs will
change from metallic to semiconducting due to Anderson localization.
 But a weak disorder only slightly affects the
conductance of armchair GNRs. The effect of edge disorder decreases
as the width of GNRs increases. So narrow zigzag GNRs with a weak
disorder can be used as functional elements  in a nanodevice. And
GNRs used as connections should be wider than GNRs used as
functional elements. These results are useful for better
understanding the property of realistic graphene nanoribbons, and
will be helpful for designing nanodevices based on graphene.

\vspace{0.5cm}

The authors would like to thank N. M. R. Peres for helpful
discussions.

\bibliography{GrapheneRibbons}

\begin{thebibliography}{40}
\bibitem[{eta()}]{eta}
\bibinfo{note}{We use $\eta=10^{-4}$eV for GNRs with a sigle defect, and
  $\eta=10^{-6}$eV for GNRs with a weak disorder.}
\end{thebibliography}

\end{document}